\documentclass[aps,PRMaterials,reprint,twocolumn]{revtex4-1}
\usepackage{amsmath,amssymb}
\usepackage{subfigure}
\usepackage[colorlinks,linkcolor=red,anchorcolor=green,citecolor=blue]{hyperref}
\usepackage{graphicx}
\usepackage{bm}
\usepackage{color}
\usepackage{times}
\usepackage{siunitx}
\usepackage{mathtools}
\usepackage{float}
\usepackage{tabularx}
\usepackage{ulem}

\begin{document}
\title{Directional Design of Materials Based on the Multi-Objective Optimization: A Case Study of Two-Dimensional Thermoelectric SnSe}

\author{Shenshen Yan$^{1}$}
\author{Yi Wang$^{1}$}
\author{Zhibin Gao$^{1,2}$}
\author{Yang Long$^{1}$}
\author{Jie Ren$^{1,3}$}
\email[Corresponding address: ]{Xonics@tongji.edu.cn}
\affiliation{$^{1}$Center for Phononics and Thermal Energy Science, China-EU Joint Lab on Nanophononics, Shanghai Key Laboratory of Special Artificial Microstructure Materials and Technology, School of Physics Science and Engineering, Tongji University, Shanghai 200092, China}
\affiliation{$^{2}$Department of Physics, National University of Singapore, Singapore 117551, Republic of Singapore}
\affiliation{$^{3}$Shanghai Research Institute for Intelligent Autonomous Systems, Tongji University, Shanghai 200092, China}
\date{\today}
%Phone number: 18800293990(Shenshen Yan); 18817366156(Yi Wang); 13681788319(Jie Ren)
%Email: 5shshy@tongji.edu.cn; wangyigo@tongji.edu.cn; xonics@tongji.edu.cn
\begin{abstract}
Directional design of functional materials with multi-objective constraints is a big challenge, whose performance and stability are determined by different physics factors entangled with each other complicatedly. In this work, we apply the multi-objective optimization based on the Pareto Efficiency and Particle-Swarm Optimization methods to design new functional materials directionally. As a demonstration, we achieve the thermoelectric design of 2D SnSe materials through the methods. We identify several novel metastable 2D SnSe structures with simultaneously lower free energy and better thermoelectric performance over the experimentally-reported monolayer structures. We hope our results about the multi-objective Pareto Optimization method can make a step towards the integrative design of multi-objective and multi-functional materials in the future.
\end{abstract}

\maketitle
The thermoelectric devices and materials can directly convert the electricity into thermal energy for cooling or generate electrical power from waste heat, indicating that they have a great potential to reduce the environmental pollution and provide the cleaner energy. Moreover, the Seebeck effect can generate electrical power from waste heat described by the Seebeck coefficient and thermopower, $S=-\Delta V/\Delta T$, where $\Delta V$ is the voltage potential and $\Delta T$ is the temperature difference~\cite{Bell1457}. Thermopower plays an important role in the thermoelectric materials.

Recently, bulk tin selenide (SnSe), which could undergo a second-order phase transition from the low symmetry \textit{Pnma} to high
symmetry \textit{Cmcm} phase at 810 K,
sheds a light over the field of thermoelectric materials mostly due to its ultralow thermal
conductivity and ultrahigh power factor~\cite{zhao2014ultralow,Zhao141}. This outstanding material has rapidly aroused widespread attention in various fields including the theoretical exploration~\cite{wang2015thermoelectric,bansal2016phonon,mehboudi2016structural,skelton2016anharmonicity,PhysRevMaterials.2.054603}
and the experimental study~\cite{li2013single,zhang2014two,zhao2015controlled,Chang778,wang2018defects,doi:10.1021/jacs.8b12450,doi:10.1002/aenm.201900201}. it has a great potential for studying the thermoelectric property~\cite{CHEN2018283}, which also presents a bright prospect for the development of
the novel thermoelectric materials.
Even though there is a controversy about the single-crystalline SnSe sample and measured thermal conductivity~\cite{wei2016intrinsic}, these seminal works have made a crucial step forward to the high thermoelectric performance in simple and pure bulk materials without doping or phononic crystals.
Moreover, T. Nishimura \textit{et al.}~\cite{PhysRevLett.122.226601} revealed the convergency of new Fermi pockets of the SnSe \textit{Pnma} structure at about 0.89 GPa.
On account of the quantum confinement effect, two dimensional (2D) materials have some unexpected properties compared with their pristine bulk counterparts~\cite{xu2014length,gao2017novel,shimizu2016enhanced}.

\begin{figure}
	\includegraphics[width=\columnwidth]{1.pdf}
	\vspace{-0.3cm} \caption{\label{fig:figure1}	
	(a) Three-view drawings of two typical 2D SnSe structures corresponding to monolayers of the  \textit{Pnma} and \textit{Cmcm} bulk phases. The dark gray and the green atom symbols are the Sn and Se, respectively. (b) The workflow of directional design for the 2D SnSe
materials based on the Pareto Multi-objective Optimization. }
\end{figure}

Furthermore, the materials discover and design using machine learning strategies to explore the materials design ~\cite{recent-advances-npj, PhysRevMaterials.2.120301} has made a remarkable progress. Recently, the clustering and inverse design of topological materials with machine learning have been applied in the phononic ~\cite{long2020unsupervised} and photonic~\cite{long2019inverse} materials. As the same time, applying machine learning strategies could guide to discover and design more thermoelectric materials with high performance~\cite{npj-thermalelectric-review,scientific-report-ML-TE}.
Several monolayer SnSe structures have been discovered recently~\cite{C7NR04766E,PhysRevB.97.075438,C8CP07645F}. With the innovation, it is worth exploring the more novel stable 2D SnSe materials with high thermopower. However, it is a multi-objective optimization problem to satisfy two objectives, the stability and the performance of generating electrical power from waste heat at the same time.

\begin{figure*}
	\includegraphics[width=2\columnwidth]{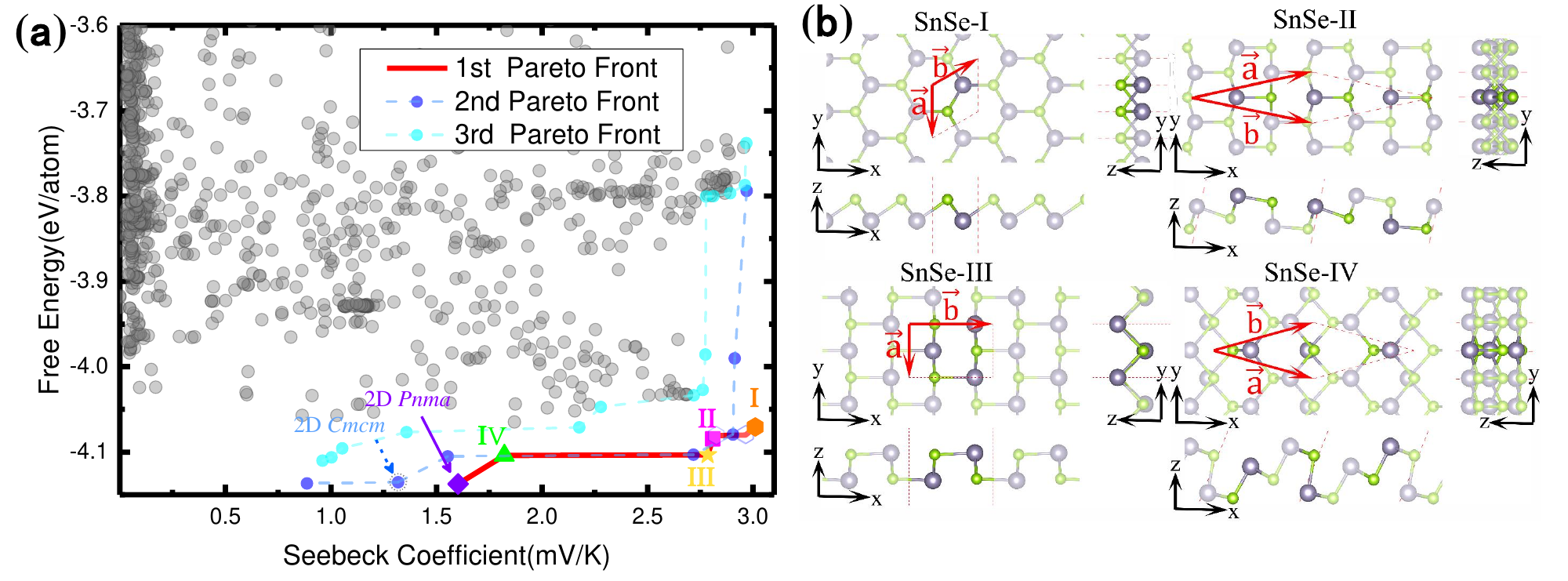}
	\vspace{-0.4cm} \caption{\label{fig:figure2}		
		The thermopower at room temperature (300 K) landscape versus free energy of 2D SnSe materials and the monolayer structures on the first Pareto Front. (a)The red line is the first Pareto Front, the larger colorful symbols represent monolayer structures with simultaneously larger Seebeck coefficient and lower free energy than others, all the circles. The light blue circles with blue dash line and the light cyan circles with cyan dash line are the second and third Pareto Fronts, respectively. The grey circles are the dominated structures by the first three Pareto Fronts. The violet rhombus and the near light blue circle with dot circle are the monolayers of the \textit{Pnma} phase and \textit{Cmcm} phase, respectively. (b) three-views of four new monolayer structures on the first Pareto Front, SnSe-\uppercase\expandafter{\romannumeral1},
SnSe-\uppercase\expandafter{\romannumeral2},
 SnSe-\uppercase\expandafter{\romannumeral3} and SnSe-\uppercase\expandafter{\romannumeral4}, corresponding to the orange hexagon, magenta square, yellow pentagram and green triangle in (a), respectively.
The blue hollow hexagons on the first Pareto Front corresponds to bilayer SnSe structures discussed in the Supporting Information~\cite{SI}.}
\end{figure*}
In the mapping from structure to function, the microscopic atomic configuration is the core of macroscopic properties.
Owing to the significant development of crystal structures prediction, new materials can be
inversely searched by artificial intelligence such as Genetic Algorithm (GA)~\cite{oganov2006crystal},
Particle Swarm Optimization (PSO)~\cite{wang2010crystal}, etc. rather
than the traditional costly Edisonaian trial-and-error approach.
The metastable structures are also important and may have high performance in some properties~\cite{RN4243, Cho625}.
Some multiple objective algorithms~\cite{NUNEZVALDEZ2018152,RN4235} have been applied to deal with the multi-objective problem efficiently.
Due to the multi-objective Particle Swarm Optimization (MOPSO) algorithm is simple to implement relatively with less hyper-parameters. Meanwhile, it also has faster convergence rate for global or local optima with sharing the information among the particles~\cite{yang2009novel,lalwani2013comprehensive,1304847}.
There are some studies to search for novel atomic SiO$_{2}$ monolayers with negative Poisson's ratio~\cite{gao2017novel} and HfO$_{2}$ monolayers with high static dielectric constant ~\cite{wang2020computational} through the multi-objective Optimization, which are combining the first-principle calculations with GA.

In this letter, we adopt the efficient and reliable algorithm, MOPSO-sigma~\cite{NUNEZVALDEZ2018152}, to directionally design 2D SnSe materials with the lower free energy and the larger thermopower based on the Pareto Optimality. In the process, The Pareto Fronts are obtained by the fast Non-dominated sorting approach~\cite{deb2002fast}.

During the design process with MOPSO algorithm, the self-consistent energy calculations and structure optimization were
employed using the Perdew-Burke-Ernzerhof (PBE) exchange-correlation functional~\cite{perdew1996}
along with the projector-augmented wave (PAW) potentials~\cite{blochl1994} implemented in the Vienna
Ab-initio Simulation Package (VASP)~\cite{kresse1996}. The  energy convergence threshold is set to 10$^{-4}$ eV
and all the atoms are allowed to relax until the maximal Hellmann-Feynman force is less than 0.001 eV/{\AA}. After this process, in order to get accurate results of crystal structure, the band structure, thermopower and phonon dispersion, we reset the energy convergence threshold and maximal Hellmann-Feynman force to the 10$^{-8}$ eV and 10$^{-6}$ eV/{\AA}.
The kinetic energy cut-off was 500 eV and phonon dispersion was obtained using Phonopy package~\cite{togo2008first}.
The Seebeck coefficient was evaluated by BoltzTraP~\cite{madsen2006boltztrap} and the MOPSO approach~\cite{nunez2018efficient,zhang2015inverse}
discussed here has been implemented in our homemade computer code.

% Describe the  in figure1.
The workflow of directional design for the thermoelectric materials based on the Pareto Optimality is shown in Fig. {\ref{fig:figure1}(b). Firstly,
the initial crystal structures are generated through the atomic Wyckoff position and 230 space groups, and similar structures are removed in order to avoid wasting computational resources~\cite{wang2010crystal}.
Then
local optimization (structure relaxation) is utilized sequentially to eliminate some
worst structures and this crucial process guarantees population diversity and makes whole energy
landscapes a well-organized shape.
Thirdly,
two objectives, free energy and thermopower, can be calculated. Here, as a benchmark, we only consider the thermopower at room temperature (300 K) data in all calculations.
The next pivotal step
is to apply multi-objective method to design the structures with the lower free energy and higher Seebeck coefficient
based on the Pareto Optimality. Different from any other single objective optimization algorithm, the leader
is not one structure with unique property, but a leader set, which is named
the Pareto Front including the structures with lower free energy and the larger thermopower.
Generally, in a collection of 2D SnSe materials, $\left\{M_n\right\}=\left\{M_1,...M_i,...M_j,...M_N \right\}$, $n\in$ $\left\{1,2,\dots, N \right\}$,
\textit{N} is the number of all structures. If $ M_i $ possesses the lower free energy and larger thermopower than $M_j$, it means that ${ M_j }$ is dominated by ${ M_i }$, ${ M_i }$ is a non-dominated solution of ${ M_j }$, as following equations: $E(M_i) \leq E(M_j)$; and $S(M_i)\geq S(M_j)$, where the $E(M_i)$ and $S(M_i)$ are free energy and thermopower of the structure $M_i$, respectively.

If ${ S_i }$ is not dominated by any other, the ${ S_i }$ is regarded as a Pareto Optimality or Pareto Efficiency structure. All the Pareto Optimality structures constitute the Pareto Front. Then we apply the PSO algorithm to make the Pareto Front lead all the population structures forward to the next Pareto Front~\cite{NUNEZVALDEZ2018152} with lower free energy and larger thermopower. Finally, if the convergence criterion is reached, the whole procedure will stop and output the reasonable Pareto Front. Otherwise, return and repeat all the above processes.
\begin{figure*}
	\includegraphics[width=2\columnwidth]{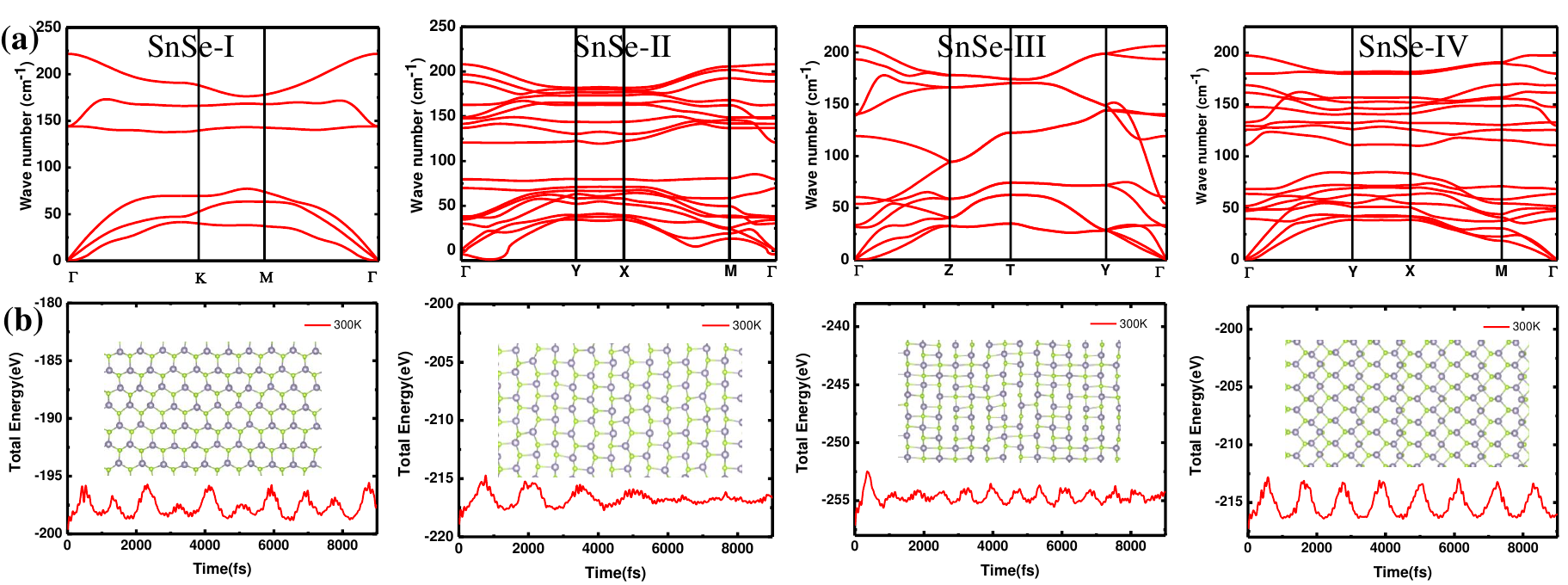}
	\vspace{-0.4cm} \caption{\label{fig:figure3}
        The phonon dispersions and AIMD simulations at 300K of SnSe-\uppercase\expandafter{\romannumeral1},
SnSe-\uppercase\expandafter{\romannumeral2}, SnSe-\uppercase\expandafter{\romannumeral3} and SnSe-\uppercase\expandafter{\romannumeral4}. (a) The phonon dispersions along the high symmetry points. There is not any imaginary frequency in the phonon dispersions of SnSe-\uppercase\expandafter{\romannumeral1}, SnSe-\uppercase\expandafter{\romannumeral3} and SnSe-\uppercase\expandafter{\romannumeral4}, and there are little imaginary frequencies near $\Gamma$ point in the phonon dispersion of SnSe-\uppercase\expandafter{\romannumeral2}. (b)The total energy varies with time during the AIMD simulations at 300K with lifetime 9 ps. The oscillating redlines present the changes of the total free energy during the AIMD simulations. The inset pictures of AIMD simulations are the top views of SnSe-\uppercase\expandafter{\romannumeral1}  ,
SnSe-\uppercase\expandafter{\romannumeral2}, SnSe-\uppercase\expandafter{\romannumeral3} and SnSe-\uppercase\expandafter{\romannumeral4} structures after the AIMD simulations, respectively.}
\end{figure*}

% Describe the result of Pareto Front set.
The result of adopting multi-objective is shown in Fig. \ref{fig:figure2}(a), which includes 2700 samples in the process to design the stable functional
structures only based on the chemical composition.
For examples, there are only first three Pareto Fronts highlighted in the Fig. {\ref{fig:figure2}}(a).
The red line is the first Pareto Front set inherently with the lower free energy and the larger thermopower than other structures presented by circles. Interestingly, we find the
known 2D SnSe structures~\cite{wang2015,zhang2016tinselenidene}, the violet rhombus symbols and its left nearest neighbor gray dash circles corresponding to 2D \textit{Pnma} and 2D \textit{Cmcm} structures of SnSe shown in Fig. \ref{fig:figure1}(a), respectively, during our blind searching processes, which does validate the correctness and exhibit robustness of our multi-objective design method.
Furthermore, as shown in Fig. \ref{fig:figure2}(b), there are four novel monolayer materials with superior stability and much larger thermopower than the known structures in Fig. \ref{fig:figure1}(a), namely, SnSe-\uppercase\expandafter{\romannumeral1}, SnSe-\uppercase\expandafter{\romannumeral2}, SnSe-\uppercase\expandafter{\romannumeral3} and SnSe-\uppercase\expandafter{\romannumeral4}, which are corresponding to the orange hexagon, magenta square, yellow pentagram and green triangle on the first Pareto Front, respectively.
 Here, we only focus on their monolayer structures and ignore the bilayer phases. Those bilayer phases are corresponding to the blue hollow hexagons on the first Pareto Front, as shown in the Supporting Information.~\cite{SI}. Moreover, the light blue circles with blue dash line and the light cyan circles with cyan dash line are the second and third Pareto Fronts, respectively. The monolayer of Cmcm phase is also found corresponding the light blue circle with dot circle lied on the second Pareto Front. More structures on the second and the third Pareto Front are shown in the Supporting Information~\cite{SI}, which contain quasi-one dimensional, bilayer and novel metastable atomic configurations. The grey circles represent the dominated structures with much higher free energy and lower thermopower than the first three Pareto Fronts, which are not discussed in this manuscript.

%\textcolor{red}{reference}
% Describe three novel crystal structures.
The structure of orange hexagon, SnSe-\uppercase\expandafter{\romannumeral1}, is
the honeycomb monolayer, similar to the silicene~\cite{cahangirov2009two} with small buckling from one side
view. The blue hollow hexagons are the bilayer counterparts
of the honeycomb structures. SnSe-\uppercase\expandafter{\romannumeral3}, the yellow pentagram, is the structure with armchair and zigzag ridges from the different side view. As well as,
we discovered accidentally that same main group and stoichiometric number of GeSe has been synthesised by high-pressure techniques~\cite{von2017high}, whose one layer is very similar to the SnSe-\uppercase\expandafter{\romannumeral3} structure.
%The structure of gray dash circle, near neighbor (left one) of the violet rhombus symbol, has higher symmetry than 2D SnSe-\uppercase\expandafter{\romannumeral2}-$\alpha$ corresponding to the monolayer \textit{Cmcm} phase compared with monolayer \textit{Pnma} phase.
The structure of magenta square, SnSe-\uppercase\expandafter{\romannumeral2}, is a combination of cells of SnSe-\uppercase\expandafter{\romannumeral1} and SnSe-\uppercase\expandafter{\romannumeral3}. More interestingly, we also find another structure, the green triangle, SnSe-\uppercase\expandafter{\romannumeral4}, is constructed with the cells of the monolayer of \textit{Pnma} and SnSe-\uppercase\expandafter{\romannumeral1}. The optimized structural properties of the novel 2D SnSe structures are summarized in the Table~\ref{tab1}.

\begin{table*}[]
\caption{Properties of the four 2D SnSe structures. The $a$ and $b$ are the lattice parameters. The $E$,  $m_{DOS,F}^{*}$, $m_{band}^{*}$, $N_v$ and $S$ are free energy per atom, density of states effective mass at the Fermi energy level, isotropic parabolic band effective mass, band degeneracy and thermopower, respectively, where the $m_{e}$ is the free electron mass.}
\setlength{\tabcolsep}{3mm}{
\begin{tabular}{llccccccc}
\toprule
\multicolumn{1}{c}{Materials} &Space group (NO.) & $a$ (\AA)    & $b$ (\AA)    & $E$ (eV/atom) & $m_{band}^{*}$ ($m_{e}$) & $m_{DOS,F}^{*}$ ($m_{e}$)  & $N_v$ & $S$ (mV/K) \\ \hline
SnSe-I                        &\textit{P3m1} (156) & 3.91 & 3.91 & -4.08 & 1.34 & 8.03          & 6  & 3.587   \\
SnSe-II                       &\textit{Cm} (8) & 9.64 & 9.64 & -4.09 & 2.19 & 4.38           & 2  & 2.763   \\
SnSe-III                      &\textit{Pmmn} (59) & 3.92 & 6.17 & -4.10 & 1.31 & 2.62           & 2  & 2.596   \\
SnSe-IV                       &\textit{Cm} (8) & 7.67 & 7.67 & -4.11 & 0.37, 0.76 & 1.13     & 2  & 1.871   \\ \hline
\botrule
\end{tabular}}
\label{tab1}
\end{table*}

Then we confirm the thermoelectric performance and the stabilities of the novel 2D SnSe structures on the first Pareto Front adopting the multi-objective method, which have light higher free energy and much larger thermopower than the structures in Fig. {\ref{fig:figure1}(a).
The first priority is to certify the stabilities of these new freestanding 2D SnSe materials.
The free energy of SnSe-\uppercase\expandafter{\romannumeral1} (-4.079 eV/atom),
SnSe-\uppercase\expandafter{\romannumeral2} (-4.091 eV/atom)
SnSe-\uppercase\expandafter{\romannumeral3} (-4.103 eV/atom) and  SnSe-\uppercase\expandafter{\romannumeral4} (-4.106 eV/atom)
are higher than free energy of \textit{Pnma} phase (-4.137 eV/atom). The dynamical stabilities of these metastable structures have been confirmed by phonon dispersion relations at 0 K and ab initio molecular dynamics (AIMD) simulations at 300 K shown in Fig. {\ref{fig:figure3}}(a) and Fig. {\ref{fig:figure3}}(b). There is not any imaginary frequency in the phonon dispersions of the 2D SnSe-\uppercase\expandafter{\romannumeral1}, SnSe-\uppercase\expandafter{\romannumeral3} and SnSe-\uppercase\expandafter{\romannumeral4}.
And there are little imaginary frequencies near $\Gamma$ point in the phonon dispersion of 2D SnSe-\uppercase\expandafter{\romannumeral2}.
For 2D materials, %there are two acoustic phonon modes linear with \textbf{q} in the vicinity of the Brillouin Zone center and the cross-plane polarized acoustic ZA mode has a quadratic dispersion near the $\Gamma$ point that is a characteristic of 2D materials.
the imaginary frequencies near the $\Gamma$ point were also reported in germanene \cite{cahangirov2009two}, honeycomb arsenenes \cite{kamal2015arsenene} and this artificial imprecision in the computation has nothing to do with a structural transition and instability \cite{carrete2016physically,liu2016continuum,csahin2009monolayer}.
After AIMD with the NVT ensemble simulations, the atomic configurations are changed to the position of equilibrium at 300K.
Furthermore, the AIMD simulations with NVE ensemble at a series of elevated temperatures with lifetime 9 ps at 300 K are performed to verify their dynamic stabilities with the oscillating free energy shown in Fig. {\ref{fig:figure3}}(b). The top views of 2D SnSe-\uppercase\expandafter{\romannumeral1}, SnSe-\uppercase\expandafter{\romannumeral2}, SnSe-\uppercase\expandafter{\romannumeral3} and SnSe-\uppercase\expandafter{\romannumeral4} after the AIMD simulations are shown in the inset pictures. They show the dynamic stabilities of the four structures.

\begin{figure*}
	\includegraphics[width=2\columnwidth]{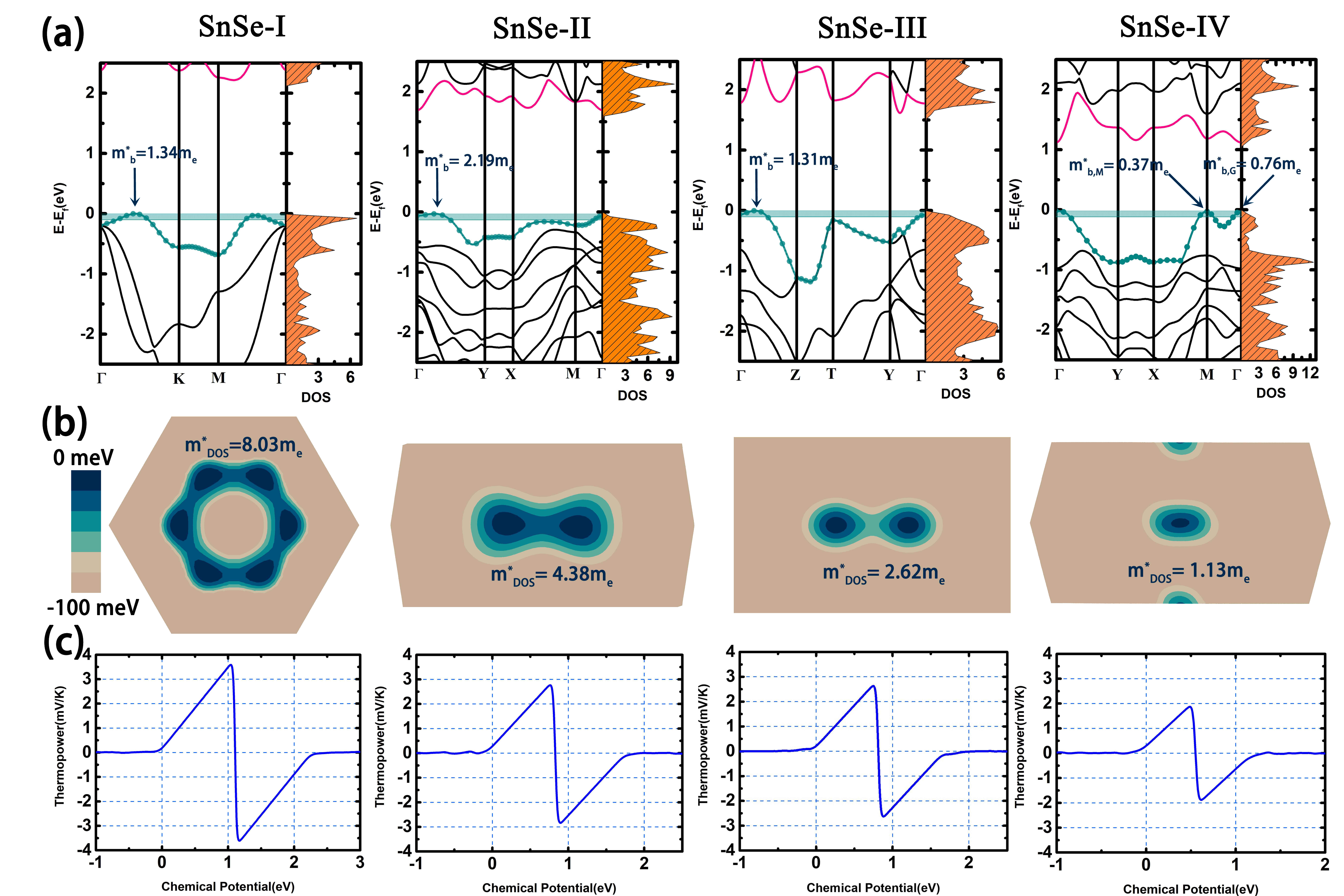}
	\vspace{-0.3cm} \caption{\label{fig:figure4}
Electronic band structure linked density of states(DOS), Fermi surface and thermopower of SnSe-\uppercase\expandafter{\romannumeral1}, \uppercase\expandafter{\romannumeral2}, \uppercase\expandafter{\romannumeral3}, \uppercase\expandafter{\romannumeral4}.
(a) Electronic band structure linked DOS. the cyan line and symbol are the valance band falling on the 100 meV energy window of the VCM, which are the cyan flat band. The Fermi energy is set as 0 ev.
(b) Fermi surface in the first Brillouin zone in the 100 meV energy window of the VCM corresponding to the cyan flat band in the (a).
(c) Average Seebeck coefficient versus Chemical potential.}
\end{figure*}

Electronic band structure linked density of states(DOS), Fermi surface and thermopower of SnSe-\uppercase\expandafter{\romannumeral1}, SnSe-\uppercase\expandafter{\romannumeral2}, SnSe-\uppercase\expandafter{\romannumeral3} and SnSe-\uppercase\expandafter{\romannumeral4},
are shown in Fig. \ref{fig:figure4}. We can find that in the structure SnSe-\uppercase\expandafter{\romannumeral1}, the valance band maximum (VBM) is in the $\Gamma$-K path, and the conduction band minimum (CBM) is in the M-$\Gamma$ path in the Brillouin Zone (BZ) with an indirect band gap of 2.22 eV.
In the SnSe-\uppercase\expandafter{\romannumeral2}, the VBM is in the Y-$\Gamma$ path, and the CBM is in the $\Gamma$ point with an indirect band gap of 1.66 eV.
For the SnSe-\uppercase\expandafter{\romannumeral3}, the VBM is in the Y-$\Gamma$ path, the CBM is in the $\Gamma$-Z path with the indirect band gap of 1.61 eV.
Furthermore, in the SnSe-\uppercase\expandafter{\romannumeral4}, both VBM and CBM are in the $\Gamma$ point with a direct band gap of 1.10 eV. Their band gaps are larger than that of the monolayer \textit{Pnma} phase (0.91 eV~\cite{wang2015thermoelectric}).
In the four structures, the sharp peaks of DOS near the valence maxima and conduction minima may enhance the thermopower.

Figure. \ref{fig:figure4}(c) shows the average thermopower depending on the chemical potential. The thermopower along the x and y directions is shown in the Supporting Information~\cite{SI}. The positive and negative signs of thermopower correspond to the hole carrier (p-type and $\mu <0$) and electron carrier (n-type and $\mu >0$). The maximum of thermopower with p-type carrier concentrations of SnSe-\uppercase\expandafter{\romannumeral1}, SnSe-\uppercase\expandafter{\romannumeral2}, SnSe-\uppercase\expandafter{\romannumeral3} and SnSe-\uppercase\expandafter{\romannumeral4} are 3.587 ($mV/K$), 2.763($mV/K$), 2.596 ($mV/K$) and 1.871 ($mV/K$) which are much larger than the known 2D SnSe monolayer of \textit{Pnma} phase (around 1.75 ($mV/K$))~\cite{shafique2017thermoelectric}
and 3-6 times of bulk SnSe (0.5-0.58 ($mV/K$))~\cite{dewandre2016two}.

Generally speaking, there are several approaches to enhance the thermopower, including
high valley degeneracy produced by carrier pocket engineering~\cite{pei2011convergence,PhysRevLett.108.166601,wang2018defects,PhysRevMaterials.2.054603,doi:10.1021/jacs.8b12450,PhysRevLett.122.226601}, a distorted DOS by doping that resonates one energy level of a localized atom~\cite{heremans2008}, weak electron-phonon coupling~\cite{wang2012weak}, phonon drag effect~\cite{ohta2007giant} and pudding-mold-like shape in the highest valence band or lowest conduction band that is beneficial to a high Seebeck and
conductivity~\cite{kuroki2007pudding}. In the Mott formula~\cite{mahan1996best}, $S= \frac{\pi^2 k_{B}^2 T} {3q} \left.\frac{d[\ln(\sigma(E))] } {dE} \right|_{E=E_F}$, where $\sigma(E)=g(E)f(E)q \mu$ is the energy-dependent electrical conductivity. Meanwhile, $\sigma(E)$ is dominated by $f(E)$, $g(E)$, $q$, $k_{B}$, and $\mu$, the Fermi-Dirac distribution function, the DOS, the carrier charge, the Boltzmann constant and the mobility, respectively. Generally, the materials have a local substantially increase of the energy-dependent DOS at the Fermi energy, the thermopower will be significantly enhanced.

For the 2D materials, the correlation between the thermopower and the DOS effective mass at the Fermi energycan be expressed as following equation (details can be found in the Appendix): different from thermopower of 3D materials $S_{3D}= \frac{2\pi^2 k_{B}^2 T} {3q\hbar^{2}}(\frac{1}{3\pi^{2}n})^{2/3}m^{*}_{DOS,F}$,
%\begin{equation}\label{seebeck2}
 for the 2D materials, $S_{2D}= \frac{\pi k_{B}^2 T} {6qn\hbar^2}  m_{DOS,F}^*$,
where $\hbar$ is reduced Planck constant, $m_{DOS,F}^*$ is the DOS effective mass at the Fermi energy level, $n$ is the carrier concentration. Therefore, the larger DOS effective mass will lead to a larger thermopower.

%band degeneracy
For the 2D materials, the correlation between the DOS effective mass ($m_{DOS}^*$) and the isotropic parabolic band effective mass ($m_{band}^{*}$) is:
\begin{equation}
m_{DOS}^{*}=N_{v}m_{band}^{*}
\end{equation}
different from the 3D density of
states effective mass formula: $m_{DOS} = N_{v}^{2/3} m_{band}$, where the $N_{v} $  is the band degeneracy (details can be found in the Appendix). In the sense of p-type doping, the Fermi energy is close to the top of the valence band, and the Fermi surfaces within 100 meV from the top of the valence band in the first Brillouin zone are  calculated, as shown in the  Fig. \ref{fig:figure4}(b). From the electronic band structures and Fermi surfaces of the SnSe-\uppercase\expandafter{\romannumeral1},  SnSe-\uppercase\expandafter{\romannumeral2}, SnSe-\uppercase\expandafter{\romannumeral3} and SnSe-\uppercase\expandafter{\romannumeral4} structures, we can find that the band degeneracy, $N_{v}$, are 6, 2, 2, 2, and the DOS effective mass (the isotropic parabolic band effective mass) are 8.03$m_{e}$ (1.34$m_{e}$), 4.38$m_{e}$ (2.19$m_{e}$), 2.62$m_{e}$ (1.31$m_{e}$), 1.13$m_{e}$ (0.37$m_{e}$ at M point and 0.76$m_{e}$ at G point) respectively, where the $m_{e}$ is free electron mass.

According to the Mott formula and the DOS effective mass of 2D materials, we can obtain directly the thermopower for 2D materials:
\begin{equation}
 S_{2D}= N_{v}m_{band}^{*}\frac{\pi k_{B}^2 T} {6qn\hbar^2}
\end{equation}
which is different from the thermopower for 3D materials: $S_{3D}=N_{v}^{2/3}m_{band}^{*}\frac{2\pi^2 k_{B}^2 T} {3q\hbar^{2}} (\frac{1}{3\pi^{2}n})^{2/3}$. The thermopower is determined by band degeneracy($N_{v}$), band effective mass($m_{band}^*$) and carrier concentration($n$).
In addition, the band degeneracy in 2D material is more important than 3D material to the thermopower.
 For the four 2D SnSe materials, the higher band degeneracy and band effective mass could lead to the larger density of states effective mass, $m_{DOS}^{*}$, and also the higher thermopower in Table \ref{tab1}. Due to the different correlations between thermopower and band degeneracy, $S_{2D} \propto {N_{v}}$ for 2D materials and $S_{3D} \propto {N_{v}^{2/3}}$ for 3D materials, the band degeneracy, $N_v$, plays a more vital role for the 2D materials. Moreover, the band engineering strategies~\cite{doi:10.1002/adma.201202919,pei2011convergence, ding2015band, yan2015material} has ability to increase the band degeneracy for the 3D materials, which will be more effective for the 2D materials.

%{\color{red}The conclusion appears too suddenly. PLEASE add more discussions about the underlying physics, more sentences, and more perspectives about the present study before the end.}

In conclusion, we have achieved the directional design of materials through the Multi-Objective Pareto Optimization method based on the Pareto Efficiency and Particle-Swarm Optimization only according to the chemical composition. This method can design the structures with the lower free energy and the larger thermopower at the same time.
The designed novel 2D SnSe monolayers on the first Pareto Front also indicate that the main group IV-VI, for instance, the monolayer of $\beta$-GeSe at the high-pressure~\cite{von2017high} and main group-V, like abundant Phosphorene ~\cite{guan2014phase,li2017direct,Han2017Prediction} may share similar homogeneous configurations because of the similar outer valence electrons. Hence, the efficiency of Pareto Optimization of structures has demonstrated that it is instructive to the further materials design and even the experimental synthesis.

So far, we only focus on the stability and the thermopower, moreover, we explain the band degeneracy is more important to the themopower for 2D materials than 3D materials. In the future, we will add more functional objectives, such as, electronic relaxation time and lattice thermal conductivity to design the efficient thermoelectric materials, even to design the multiple functional materials.
\section*{Acknowledgements}
We acknowledge the support from the National Natural Science Foundation of China (No. 11935010 and No.11775159), the Shanghai Science and Technology Committee (Grants No. 18ZR1442800 and No. 18JC1410900), and the Opening Project of Shanghai Key Laboratory of Special Artificial Microstructure Materials and Technology.

\begin{appendix}
\section{Correlation between band effective mass and density of states effective mass, and Seebeck Coefficient for 3D and 2D materials}

  Here, we show the different correlations between Seebeck coefficient and density of states (DOS) effective mass ($m_{DOS}^*$) for 3D and 2D materials. Meanwhile, we can also obtain the correlations between band effective mass ($m_{band}^*$) and density of states (DOS) effective mass ($m_{DOS}^*$) for 3D and 2D materials.

  According to the Onsager reciprocal relations of the charge and heat currents ~\cite{BoltzTraP2}: $\label{Onsager}j_{q}= L^{(0)}\cdot\textbf{E} + \frac{L^{(1)}}{qT}\cdot(-\nabla T)$, $j_{Q}= \frac{L^{(1)}}{q}\cdot\textbf{E} + \frac{L^{(2)}}{q^{2}T}\cdot(-\nabla T)$,
where
 $L^{(\alpha)} = q^{2}\int\sigma(E)(E-E_{F})^{\alpha}(-\frac{ \partial f(E)}{ \partial{E}})dE$, we can obtain the electronic conductivity and the Seebeck coefficient under the parabolic band model approximation:
 $\label{sigma}
 \sigma_{0} = L^{(0)},
\label{S L}
 S = \frac{1}{qT} \frac{L^{(1)}}{L^{(0)}}
 $,
where the $q$, \textbf{E}, $E$, $T$, $E_{F}$ and $f(E)$ are the charge , electric field, energy, temperature, Fermi energy and  Fermi-Dirac distribution function, respectively.

When only considering the electronic transport near the Fermi level, after performing a Sommerfeld expansion, we can obtain the Mott's relations~\cite{mahan1996best}:
 $\label{Mott relations}
 S= \frac{\pi^2 k_{B}^2 T} {3q}\frac{d[ln(\sigma(E))]}{dE}\bigg|_{E=E_F}
 $.
We can find that the DOS will make an important effect on the Seebeck coefficient.

  The $\sigma(E)$ is the energy-dependent electronic conductivity, expressed as:
 $
 \sigma(E) = \tau(E)v^{2}(E)G(E)
 $,
 where the $\tau(E)$, $v(E)$ and $G(E)$ are the electronic relaxation time, group velocity of carriers and DOS of multiple degenerate bands.
 The DOS of multiple degenerate bands, $G(E)$ is determined by the DOS of single band:
  $G(E) = N_{v}g(E)$, where the $N_{v}$ and $g(E)$ are band degeneracy and the DOS of single band respectively. If only considering the acoustic phonon scattering, we can get: $\tau(E) \propto E^{-1/2}$ \cite{young2000direct}. According to the free electron approximation, the correlation between group velocity of carriers and energy is $v({E}) \propto E^{1/2}$.
 Here we will give the DOS expression for 3D and 2D systems.

For 3D system, the DOS of single band model is:
 $
 g_{3D}(E) = \frac{1}{2\pi^{2}}(\frac{2m_{band}^{*}}{\hbar^{2}})^{3/2} E^{1/2}
 $ where $\hbar$ is reduced Planck constant.
 If considering the multiple degenerate bands model, the DOS of multiple degenerate bands model, $G_{3D}(E)$ is:
 $
 G_{3D}(E) =\frac{1}{2\pi^{2}}(\frac{2m_{DOS}^{*}}{\hbar^{2}})^{3/2}E^{1/2}= N_{v}g_{3D}(E)
$.
 Then we can obtain that:
 $
 G_{3D}(E)= \frac{2}{2\pi^{2}}(\frac{2N_{v}^{2/3}m_{band}^{*}}{\hbar^{2}})^{3/2}E^{1/2}
 $.
 Therefore, we can find the correlation among the DOS effective mass, band degeneracy and single band effective mass for the 3D system\cite{doi:10.1021/jacs.6b04181, Gibbs2017Effective}:
 $
 m_{DOS}^{*} = N_{v}^{2/3} m_{band}^{*}
 $.

 Similarly, for 2D system, the DOS of single band model is:
 $
 g_{2D}(E) = \frac{2}{(2\pi)^{2}}\frac{2\pi}{\hbar} m_{band}^{*}
 $.
 If considering the multiple degenerate bands model, the DOS of multiple degenerate bands model, $G_{2D}(E)$ is:
 $
 G_{2D}(E) =\frac{2}{(2\pi)^{2}}\frac{2\pi}{\hbar} m_{DOS}^{*}= N_{v} g_{2D}(E) %\frac{2}{(2\pi^{2})} \frac{2\pi}{\hbar}N_{v} m_{b}^{*}
 $.
 Hence, we also can obtain that:
 $
 G_{2D}(E)=\frac{2}{(2\pi^{2})} \frac{2\pi}{\hbar}N_{v} m_{band}^{*}
 $.
 We can also find the correlation among the DOS effective mass, band degeneracy and single band effective mass for the 2D system:
 $
 m_{DOS}^{*} = N_{v} m_{band}^{*}
 $.
Then, we give the formulas of the Seebeck coefficient for 3D and 2D systems.

For 3D system, the energy-dependent electronic conductivity: $\sigma_{3D}(E)\propto E$, we could obtain the Seebeck coefficient:

 \begin{equation}
 \begin{split}
 S_{3D} &= \frac{\pi^2 k_{B}^2 T} {3q} \frac{d[ln(\sigma(E))] } {dE}\bigg|_{E=E_F} \\
 &= \frac{\pi^2 k_{B}^2 T} {3q} \frac{1}{E_{F}}\\
 &=\frac{\pi^2 k_{B}^2 T} {3q}\frac{2m_{DOS,F}^{*}}{\hbar^{2}k_{3D,F}^{2}}\\
 &= \frac{2\pi^2 k_{B}^2 T} {3q\hbar^{2}}(\frac{1}{3\pi^{2}n})^{2/3}m^{*}_{DOS,F}
 \end{split}
 \end{equation}
 where $m_{DOS,F}^{*}=\frac{\hbar^{2}k_{F}^{2}}{2E_{F}}$ is the DOS effective mass at the Fermi energy level, $k_{F}$ is the Fermi radius, the $k_{3D,F} = (3\pi^{2}n)^{1/3}$ is the Fermi radius for 3D system, and $n$ is carrier concentration.

Similarly, for 2D system, the energy-dependent electronic conductivity: $\sigma_{2D}(E)\propto E^{1/2}$, we could obtain the Seebeck coefficient:

\begin{equation}
  \begin{split}
 S_{2D} &= \frac{\pi^2 k_{B}^2 T} {3q}\frac{d[ln(\sigma(E))] } {dE}\bigg|_{E=E_F} \\
 &= \frac{\pi^2 k_{B}^2 T} {3q}  \frac{1}{2} \frac{1}{E_{F}}\\
 &= \frac{\pi^2 k_{B}^2 T} {3q} \frac{1}{2}\frac{2m_{DOS,F}^{*}}{\hbar^{2}k_{2D,F}^{2}}\\
 &=\frac{\pi k_{B}^2 T}{6qn\hbar^2}m^{*}_{DOS,F}
 %&=\frac{2\pi^3 k_{B}^2 T}{3qh^2}m_{DOS,F}^*(\frac{1}{n}), \\
 \end{split}
 \end{equation}
 where the $k_{2D,F}=(2\pi n)^{1/2}$ is the Fermi radius for 2D system.
\end{appendix}

\begin{widetext}

\section{Supporting Information: Directional Design of Materials Based on Multi-Objective Optimization: A Case Study of Two-Dimensional Thermoelectric SnSe}

\subsection{The thermopower along x and y directions}
\begin{figure}[H]
\includegraphics[width=\columnwidth]{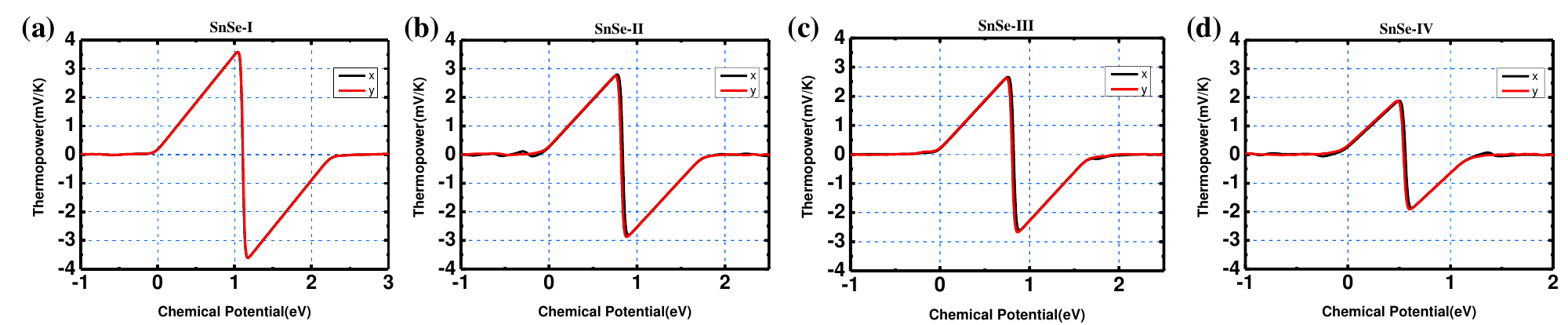}
\vspace{-0.3cm} \caption{\label{fig:figure1}
There is thermopower of the four structures along x(black line) and y(red line) directions at 300K.
}
\end{figure}
\section{The bilayer structures on the first Pareto Front}

\begin{figure}[H]
\includegraphics[width=\columnwidth]{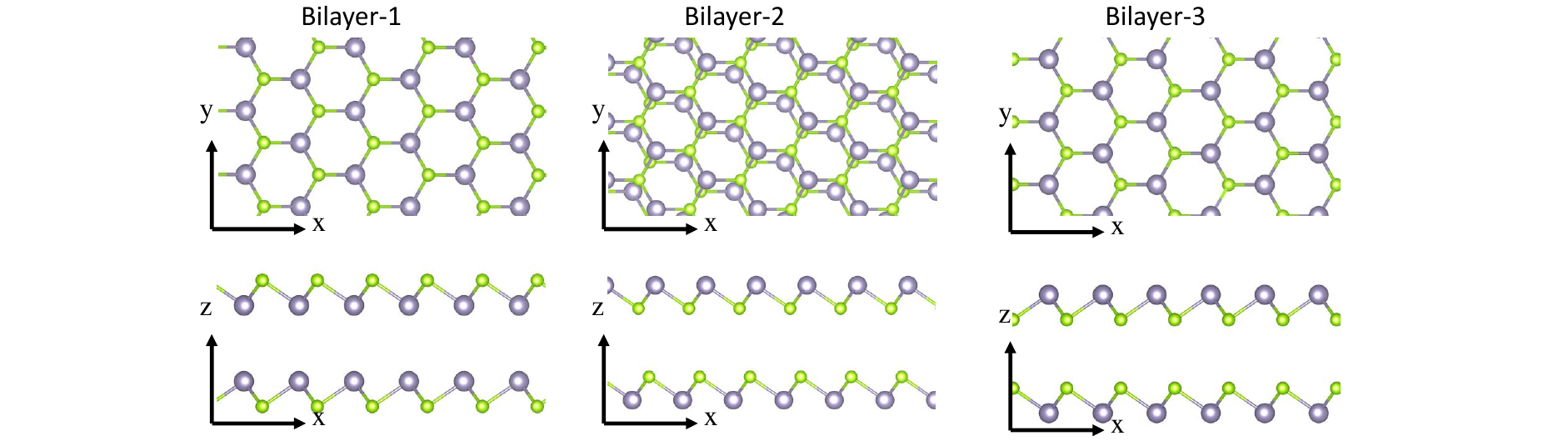}
\vspace{-0.3cm} \caption{\label{fig:figure2}
There are three bilayer structures of 2D SnSe with the top and side views, which are blue hollow hexagonal symbols, as shown on the first Pareto Front.
}
\end{figure}

\begin{table*}[h]
\caption{Properties of three bilayer structures of 2D SnSe. There are space group, the values of lattice parameters a and b,  free energy per atom and the values of thermopower, respectively.}
\setlength{\tabcolsep}{6.5mm}{
\begin{tabular}{lccccccc}
\toprule
\multicolumn{1}{c}{Materials} &Space group  & $a$(\AA)& $b$(\AA) & $E$(eV/atom) & $S$(mV/K) \\ \hline
Bilayer-1                      & P-6m2 & 3.920 & 3.920 &-4.081     &2.829           \\
Bilayer-2                      & P-1 & 3.913 & 3.915 &-4.080     &2.965           \\
Bilayer-3                      & P-6m2 &3.910 &3.910 &-4.079     &2.968           \\ \hline
\botrule
\end{tabular}}
\label{tab1}
\end{table*}

\subsection{Structures on the second Pareto Front and third Pareto Front}

\begin{figure}[H]
	\includegraphics[width=\columnwidth]{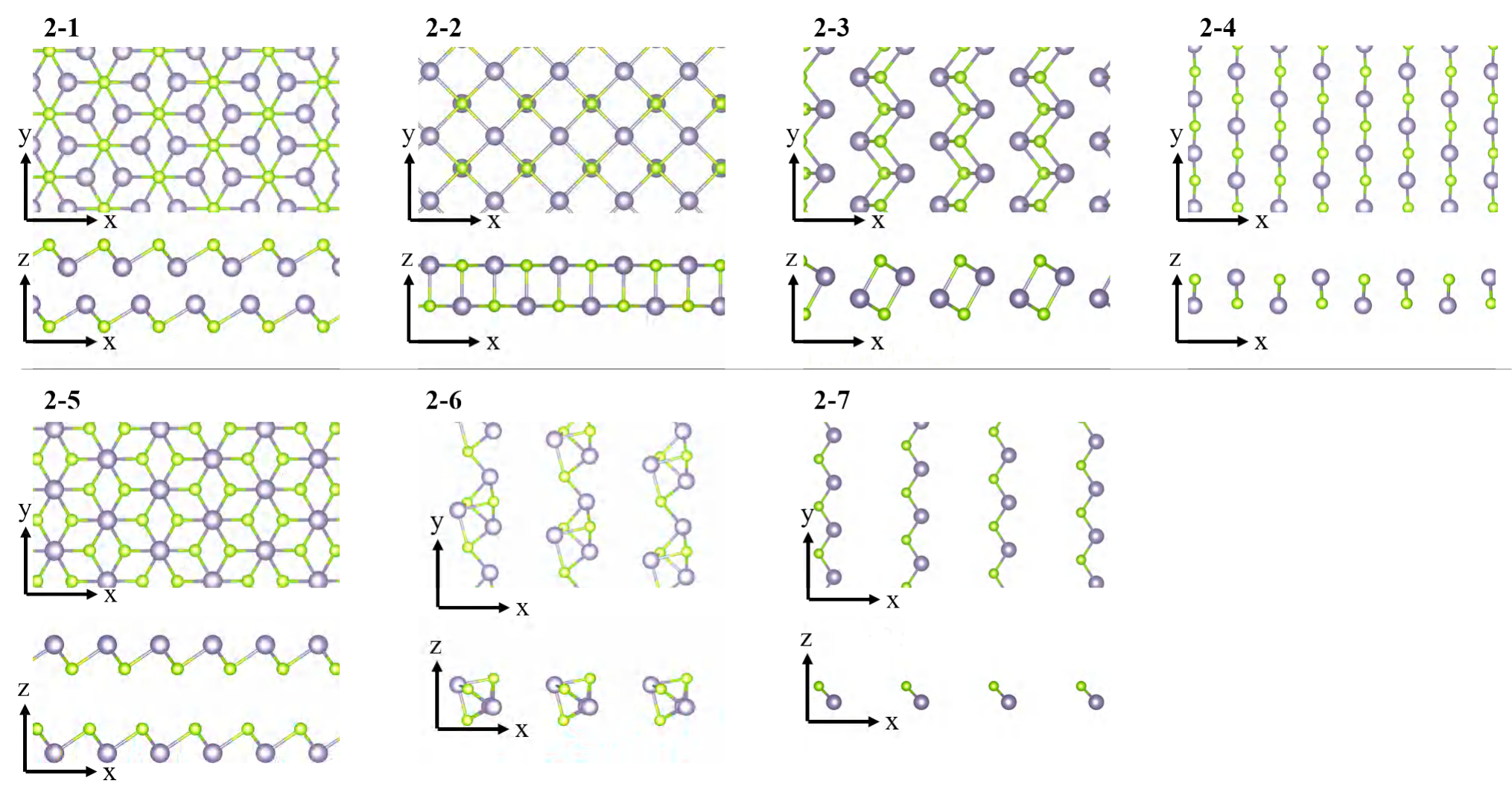}
	\vspace{-0.3cm} \caption{\label{fig:figure3}	
	Structures of 2D SnSe on the second Pareto Front are sorted by the free energy.}
\end{figure}

\begin{table*}[h]
\caption{Properties of 2D SnSe structures on the second Pareto Front. There are space group, the values of lattice parameters a and b,  free energy per atom and the values of thermopower, respectively.}
\setlength{\tabcolsep}{6.5mm}{
\begin{tabular}{lccccccc}
\toprule
\multicolumn{1}{c}{Materials} &Space group  & $a$(\AA)& $b$(\AA) & $E$(eV/atom) & $S$(mV/K) \\ \hline
2-1                      & P-3m1 &4.094 &4.094 &-4.137     &0.887           \\
2-2                      & Cmcm & 4.320 &4.320 &-4.135     &1.320           \\
2-3                      & P-1 &4.112 &5.109 &-4.105     &1.555           \\
2-4                      & Ama2 & 3.937 & 6.148 &-4.103     &2.719           \\
2-5                      & C2/m & 6.795 & 3.918 &-4.080     &2.906           \\
2-6                      & P1 & 7.031 & 7.329 &-3.990     &2.914           \\
2-7                      & P1 & 3.800 & 7.134 &-3.794     &2.971           \\
\hline
\botrule
\end{tabular}}
\label{tab2}
\end{table*}

\begin{figure}[H]
	\includegraphics[width=\columnwidth]{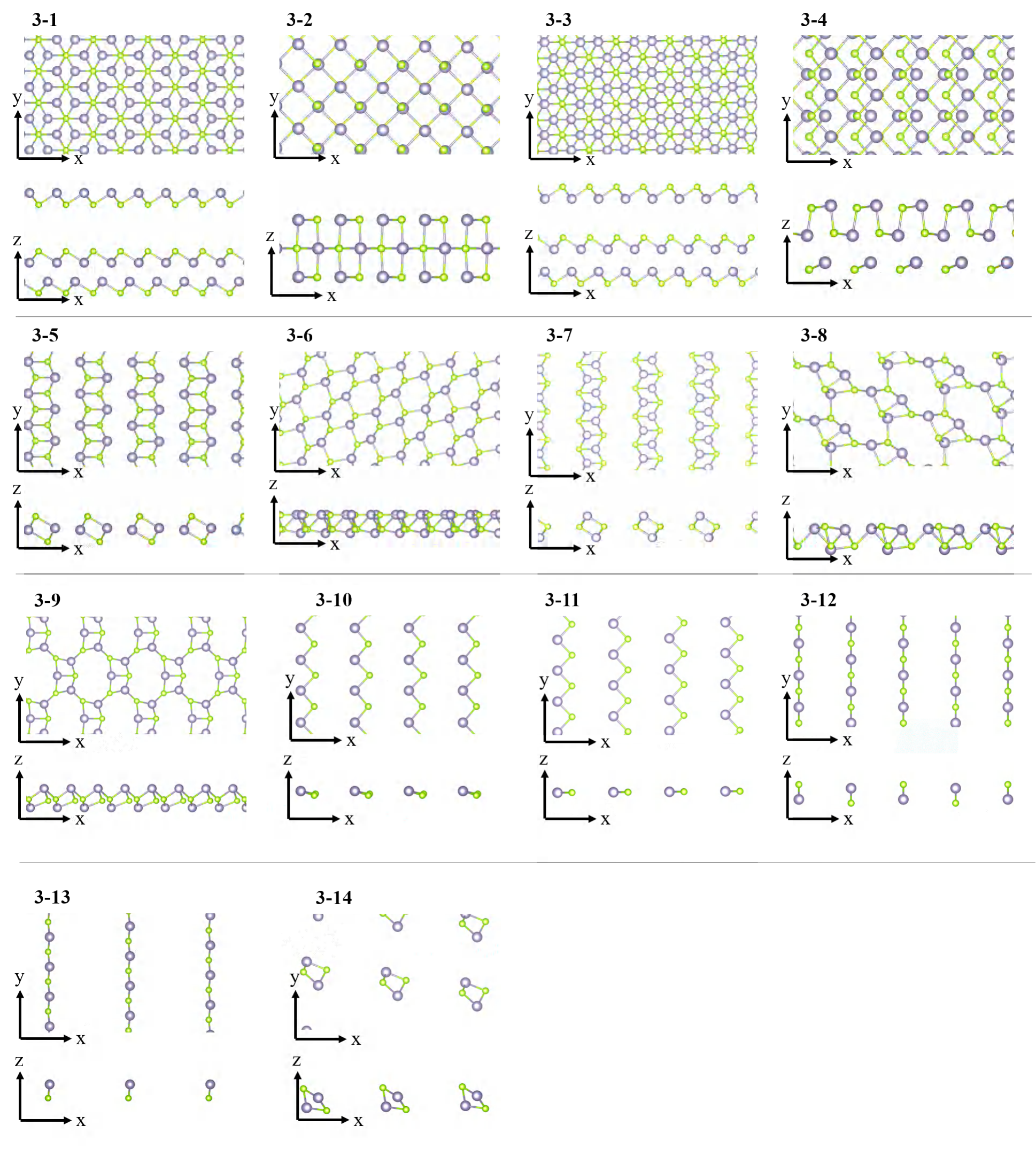}
	\vspace{-0.3cm} \caption{\label{fig:figure4}	
	Structures of 2D SnSe on the third Pareto Front are sorted by the free energy.}
\end{figure}

\begin{table*}[h]
\caption{Properties of 2D SnSe structures on the third Pareto Front. There are space group, the values of lattice parameters a and b,  free energy per atom and the values of thermopower, respectively.}
\setlength{\tabcolsep}{6.5mm}{
\begin{tabular}{lccccccc}
\toprule
\multicolumn{1}{c}{Materials} &Space group  & $a$(\AA)& $b$(\AA) & $E$(eV/atom) & $S$(mV/K) \\ \hline
3-1                      & P3m1 & 4.039 & 4.039 &-4.110     &0.961           \\
3-2                      & Cm & 6.154 & 6.039 &-4.107     &1.003           \\
3-3                      & Cm &6.825 &3.989 &-4.096     &1.054           \\
3-4                      & Pm & 4.532 & 4.134 &-4.077     &1.359           \\
3-5                      & P-1 & 3.911 & 6.410 &-4.071     &2.178           \\
3-6                      & P1 &6.516 & 6.501 &-4.047     &2.279           \\
3-7                      & P-1 & 3.955 & 7.964 &-4.034     &2.720           \\
3-8                      & P1 & 6.109 & 7.060 &-4.027     &2.762           \\
3-9                      & P1 & 7.224 & 7.232 &-3.986     &2.775           \\
3-10                      & Pm & 6.594 & 7.615 &-3.801     &2.778           \\
3-11                      &Amm2 & 3.793 & 13.768 &-3.800     &2.813           \\
3-12                     & Pmmn & 3.789 & 12.448 &-3.797     &2.894           \\
3-13                      & P1 & 3.816 & 10.770 &-3.787     &2.963           \\
3-14                      & P1 & 8.306 & 9.612 &-3.738     &2.970           \\
\hline
\botrule
\end{tabular}}
\label{tab3}
\end{table*}

\end{widetext}

\bibliography{2DSnSe-full-with-SI}
\end{document}